\documentclass[aps,twocolumn,showpacs]{revtex4}
\usepackage{amsmath,amssymb}
\usepackage{graphicx}
\usepackage{dcolumn}
\usepackage{epsf,rotate}
\usepackage[english]{babel}
\usepackage{bm}

\begin{document}

\preprint{}

%%%%%%%%%%%%%%%%%%%%%%%%%%%%%%%%%%%%%%%%%%%%%%%%%%%%%%%%%%%%%%%%%%%%%%%%%%%%%%
\title{Detecting non-linearities in data sets. Characterization of Fourier
phase maps using the Weighted Scaling Indices.}

\author{Roberto A. Monetti}
\author{Wolfram Bunk}
\author{Ferdinand Jamitzky}
\author{Christoph R\"ath}
\author{Gregor Morfill}
\affiliation{Center for Interdisciplinary Plasma
  Science (CIPS) Max-Planck-Institut f\"ur extraterrestrische
  Physik Giessenbachstr. 1, 85748 Garching, Germany}

\date{\today}

 \begin{abstract}
We present a methodology for detecting non-linearities in data sets
based on the characterization of the structural features of the Fourier
phase maps. A Fourier phase map is a $2D$ set of points $M= \{
(\phi_{\vec{k}}, \phi_{\vec{k} + \vec{\Delta}})\}$, where $
\phi_{\vec{k}}$ is the phase of the $k$-mode of the Fourier
transform of the data set and $\vec{\Delta}$ a phase
shift. 
The information thus rendered on this space is
analyzed using the spectrum of weighted scaling indices to
detect phase coupling at any scale $\vec{\Delta}$. We propose a
statistical test of significance based on the comparison of the
properties of phase maps created from both the original data
and surrogate realizations.
We have applied our method to the Lorenz system and 
the logarithmic stock returns of the Dow Jones index. Applications to higher
dimensional data are straightforward. 
The results indicate that both 
the Lorenz system and the Dow Jones time series exhibit significant signatures of
non-linear behavior. 
\end{abstract}

\pacs{05.45.-a 02.50.-r 89.20.-a}
\maketitle

%%%%%%%%%%%%%%%%%%%%%%%%%%%%%%%%%%%%%%%%%%%%%%%%%%%%%%%%%%%%%%%%%%%%%%%%%%%%%%
%\section{Introduction}
The first step in the characterization of a data set is usually the 
calculation of the 'linear
properties', i.e. the power spectrum and the amplitude
distribution. However, in many cases the estimation of higher-order
properties is required. 
For instance, given an
image it is possible to generate from it a new one by shuffling the
Fourier phases. The resulting image may look different although
the phase shuffling process preserves the power spectrum of 
the original one.  Then, the Fourier phases contain information which
is beyond the linear properties of the data, the so-called non-linear
properties. A challenging problem is then the extraction and
characterization of the information contained in the Fourier phases.
The phases
are a powerful indicator of the structure of a data set. It was early
noticed that an image synthesized by keeping the Fourier amplitudes
and changing the Fourier phases retains almost
no feature of the original image. However, image synthesized by
keeping the Fourier phases 
and changing the Fourier amplitudes conserve
most of the structure observed on the original image \cite{A}.
Studies of  
Fourier phase coupling are mostly found in the field of
astrophysics where the aim is to characterize the growth of large-scale structure in the 
universe \cite{1,2} and to test for non-Gaussian signatures in the
Cosmic Microwave Radiation Background \cite{3}. Except for these few
cases, the methods to
test for non-linearities have never focused on the analysis of phase
correlations, although some of them are based on phase randomization
procedures. This is probably due to difficulties in constructing good
estimates out of phases since they are circular quantities (defined modulo $2\pi$)
and not translational invariant, as was already noted in \cite{1,2,3}.
In this work, we present a method to analyze the Fourier phase information based 
on the assessment of the so called 'Fourier Phase Maps'. First, the method
of surrogates \cite{4,5,6,7,B} is used to generate an ensemble of
data sets which mimic the linear
properties of the original data set however wiping out higher-order correlations.
Then, we generate for both the original data set and
the surrogate data phase maps which are subsequently characterized by
means of the spectrum of weighted scaling indices \cite{15,18}. 
If the computed measure for the original data is
significantly different from the value obtained for the surrogate data 
set, one can infer that the data were generated by a non-linear
process.
% The manuscript is organized
% as follows. In section II, we introduce the Fourier phase maps. In
% section III, we present the method to asses the phase maps, namely the
% Scaling Index Method.
% Section IV contains the description of the statistical test used to
% unveil non-linearities. Section V describes applications
% of our method and we summarize our results in the last section.
%\section{The Phase Maps}

Consider an $N$-dimensional data set $\{\vec{x_i}\}$ with Fourier phases
$\{ \phi_{\vec{k}} \}$. Given a phase shift $\vec{\Delta}$ a phase map
is defined as the two dimensional set of points $M= \{ (\phi_{\vec{k}},
\phi_{\vec{k} + \vec{\Delta}}) \}$.  All possible wave numbers
$\vec{k}$ are considered up the the Nyquist frequency so as not to
include redundant information. 
The phase maps posses attracting
features which are useful to reveal non-linear properties.
First, it is well known that the Fourier transform of a random
Gaussian variable has
uncorrelated and uniformly distributed phases in
the interval $[  -\pi, \pi]$. Then, the plot of the phase map generated by such a
variable will be a uniform point distribution on the square bounded by
$y=\pm \pi$ and  $x=\pm \pi$. Here, we
generate surrogate data sets using the iterative amplitude adjusted
Fourier transform algorithm (ITAAFT) \cite{6,7}. By means of this
method, we generate surrogate data sets which keep the same 
amplitude distribution in real space and power
spectrum of the original
data set however destroying higher-order correlations. It
has been recently 
demonstrated that this algorithm can be generalized for two and three
dimensional data sets \cite{8,9}.
The surrogate method destroys higher-order properties by means of a phase
randomization procedure. Thus, we expect the
phase maps of surrogate data to be more uniform.
This is actually the
phase map feature which will be characterized in order to develop a
test for non-linearities.
%\section{Characterization of the Phase Maps: The Scaling Index Method}
%\section{Characterization of Phase Maps}

We have used the Scaling Index Method (SIM) to quantify the
structural features of the phase maps. 
This technique which was inspired in the analysis of nonlinear
systems \cite{13,14} has been successfully applied
in different fields of research, ranging from astrophysical to medical
applications \cite{15,18,9,16,17}. The SIM
characterizes the structural features of a point distribution by means of the
analysis of its local scaling behavior.

We first define
a local weighted cumulative point distribution $\rho$ as
$\rho(\vec{x}_i,R) = \sum_{j} s_R (d(\vec{x}_i,\vec{x}_j))$,
%\begin{equation}
%\label{eq1}
%  \rho(\vec{x}_i,R) = \sum_{j} s_R (d(\vec{x}_i,\vec{x}_j)) \;,
%\end{equation}
where $s_R(\bullet)$ denotes a kernel function which depends on
the scale parameter $R$ and a distance measure $d(\bullet)$.
In principle, any differentiable kernel function and any distance measure can
be used.
The weighted scaling indices $\alpha(\vec{x}_i,R)$ are then obtained by
calculating
the logarithmic derivative of $\rho(\vec{x}_i,R)$ with respect to $R$,
\begin{equation}
\label{eq2}
  \alpha(\vec{x}_i,R) = \frac{\partial \log \rho(\vec{x}_i,R)}{\partial \log R}
                      = \frac{R}{\rho}\frac{\partial}{\partial R} 
\rho(\vec{x}_i,R) \;.
\end{equation}
We use the Euclidean norm as
distance measure and a set of exponential shaping functions. So, the expression for
$\rho$ simplifies to $\rho(\vec{x}_i,R) = \sum_{j}
e^{-(\frac{d_{ij}}{R})^q}$, where $d_{ij} = \| \vec{x}_i - \vec{x}_j
\|$. 
%\begin{equation}
%\label{eq3}
%  \rho(\vec{x}_i,R) = \sum_{j} e^{-(\frac{d_{ij}}{R})^q} \;,
%                       d_{ij} = \| \vec{x}_i - \vec{x}_j \| \;.
%\end{equation}
The exponent $q$ controls the weighting of points according to their
distance to the point where $\alpha$ is calculated.
%For small values
%of $q$, points in a broad region around $\vec{x}_i$ significantly
%contribute to the weighted local density $\rho(\vec{x}_i,R)$.
%As $q$ increases, the shaping function becomes more and more
%steplike counting all points with $d_{ij} < R$ and neglecting
%all points with $d_{ij} > R$.
In this study, we use
$q=2$. The weighted scaling
indices can then be written as
\begin{equation}
\label{eq4}
  \alpha(\vec{x}_i,R) = \frac{\sum_{j} 2 (\frac{d_{ij}}{R})^2
                                             e^{-(\frac{d_{ij}}{R})^2}}
                             {\sum_{j} e^{-(\frac{d_{ij}}{R})^2}} \;.
\end{equation}
It should be noted that the use of shaping functions has two advantages, namely 
(i) one
obtains an analytical expression for $\alpha(\vec{x}_i,R)$, and (ii) the scaling
region is only determined by the parameter $R$.
The scaling-index $\alpha_i$
characterizes the structural surrounding of $\vec{x}_i$. The probability
$P(\alpha) d\alpha = \mbox{Prob}(\alpha \in [ \alpha,\alpha+d\alpha ])$
%\begin{equation}
%\label{eq5}
%  P(\alpha) d\alpha = \mbox{Prob}(\alpha \in [ \alpha,\alpha+d\alpha ])
%\end{equation}
is therefore a statistical measure of the distribution of
elementary structural components. The choice of the
scaling region is not a trivial task and is crucial to obtain meaningful
results using this technique. Although one is normally guided by the size of the
substructures under investigation, various scaling ranges should be considered
in order to obtain meaningful results.

Consider a 2D nearly uniform phase map. Since the scaling index $\alpha$ describes the
local scaling behavior, most of the $\alpha$-values
will be close to $\alpha = 2$,
leading to a nearly Gaussian frequency distribution centered
at $\alpha = 2$. However, if the phase map contains structure, the
frequency distribution will show a weaker signal around $\alpha = 2$
and new $\alpha$-values will appear. 
The circular property of the phases implies that the embedding 
square is topologically equivalent
to a torus. Then, the calculation of scaling indices was performed
using periodic boundary conditions.

We propose a statistical test of non-linearities based on the
$P(\alpha)$ spectrum. First, we generate surrogate data sets using
the ITAAFT algorithm. Then, we create for the original and the
surrogate data sets phase maps for phase shifts  $\vec{\Delta} = \{
\vec{\Delta_1}, \cdots , \vec{\Delta_m} \}$, where $\vec{\Delta_m}$ is
the maximum phase shift considered. The significance is
defined through the following expression
\begin{equation}
\label{eq6}
S(\alpha,\vec{\Delta_l}) = \frac{P_{o}(\alpha,\vec{\Delta_l})- <
  P_{s}(\alpha,\vec{\Delta_l})
  >}{\sigma(P_{s}(\alpha,\vec{\Delta_l}))} \; \; \; l=1, \cdots, m, 
\end{equation}
where $P_{o}$ is the frequency distribution of the original phase map, and  $< P_{s} >$
and $\sigma(P_{s})$ are
the mean value and the standard deviation over the frequency
distributions of the phase maps of the surrogate
realizations. Equation (\ref{eq6}) is meaningful only when 
$P_{s}(\alpha,\vec{\Delta})$  are Gaussian distributed. Although there
is no proof for this statement, simulations show that mixing processes
satisfy this condition.
Then, $S$ is measured in units of standard deviations. As mentioned
above, we expect that the original (non-linear) data set leads to
wider $P(\alpha)$ spectra while the surrogate data sets will produce
$P(\alpha)$ spectra with a stronger signal around $\alpha = 2$. Then, the statistical test
will give a positive result if the following condition is satisfied 
\begin{equation}
\label{eq7}
\alpha_i \in
\begin{cases}
\alpha: \; S(\alpha,\vec{\Delta}) < -2.6 & \bigwedge \; \; 1.90 \le \alpha \le 2.1, \\
\alpha: \; S(\alpha,\vec{\Delta}) > 2.6 & \bigwedge \; \; \alpha  \;
\; \text{elsewhere},
\end{cases}
\end{equation}
where the value 2.6 corresponds to 1$\%$ of the quantile of the normal
distribution.
However, even when the condition given by Eq. (\ref{eq7}) holds
we are not certain that original and surrogate phase maps are
significantly different since the scaling index is a
local structure measure. In fact, every scaling index value corresponds to a
region on the phase map. Thus, Eq. (\ref{eq7}) determines the region of the phase maps where
we can significantly differentiate between the original and surrogate
phase maps. In the case that this region is tiny, we can not state that the original and the
surrogate phase maps are significantly different. In order to address
this problem, we define
a new quantity $\Xi$ through the following expression
\begin{equation}
\label{eq8}
\Xi(\vec{\Delta}) = \sum_{\alpha_i} P_{o}(\alpha_i,\vec{\Delta}) \;,
\end{equation}
where $\alpha_i$ are the $\alpha$-values that satisfy Eq. (\ref{eq7}).
It should be noted that $0 \le \Xi \le 1$.
\begin{figure}
\begin{center}
\begin{minipage}{3.950cm}
     \epsfxsize=3.90cm
     \epsffile{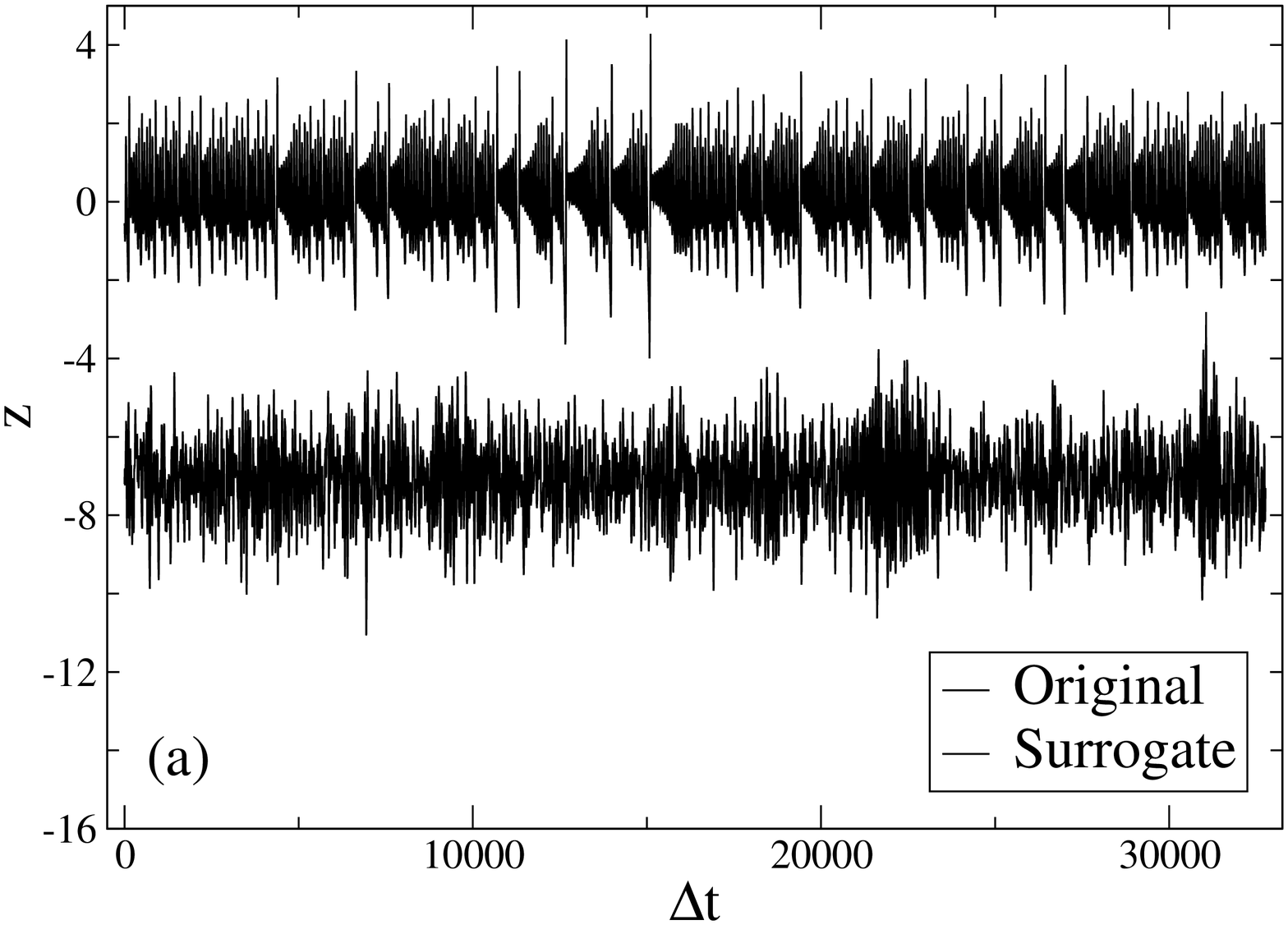}
    \end{minipage}
    \begin{minipage}{3.950cm}
    \epsfxsize=3.90cm
     \epsffile{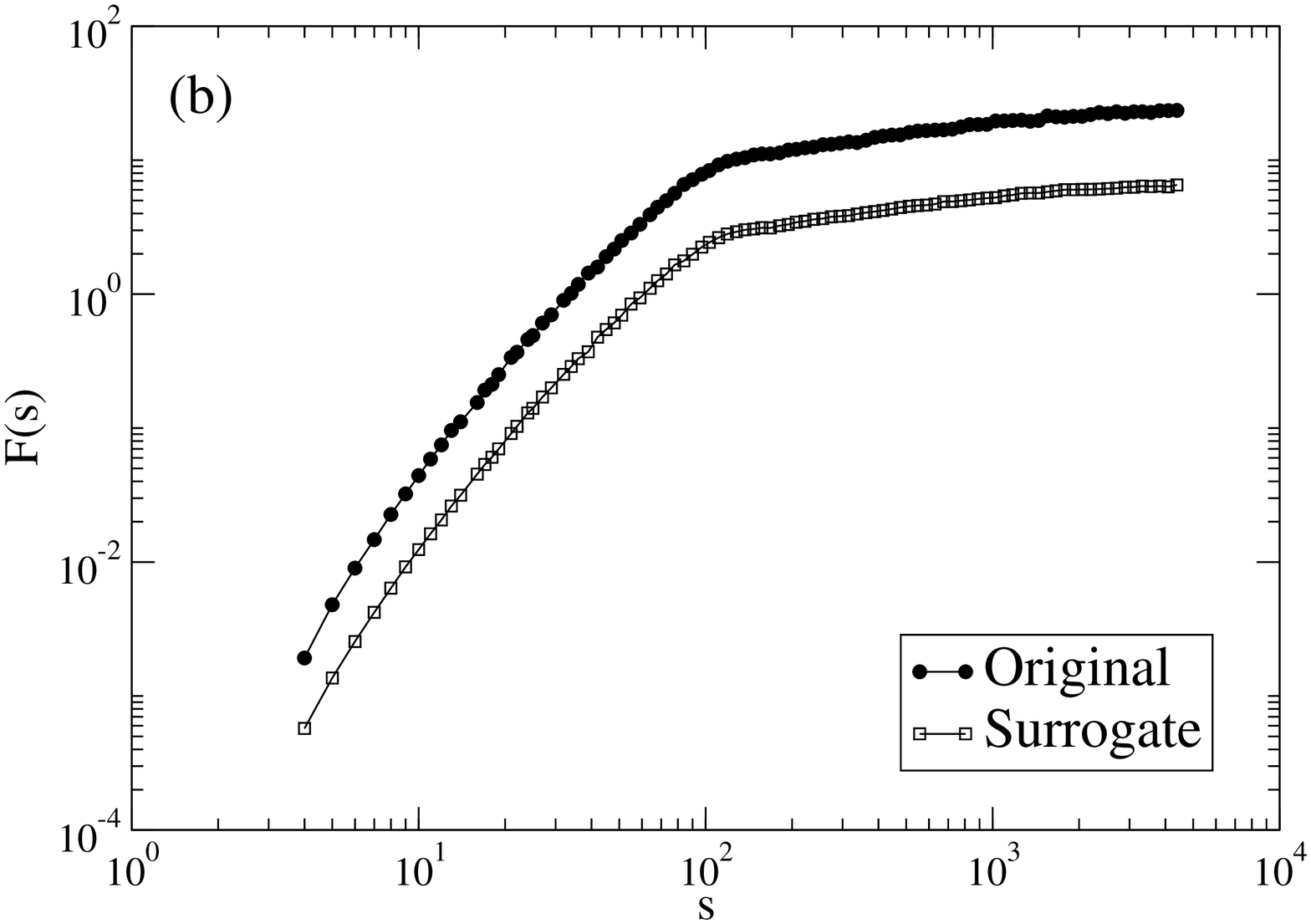}
    \end{minipage}

\begin{minipage}{4.27cm}
     \epsfxsize=4.27cm
     \epsffile{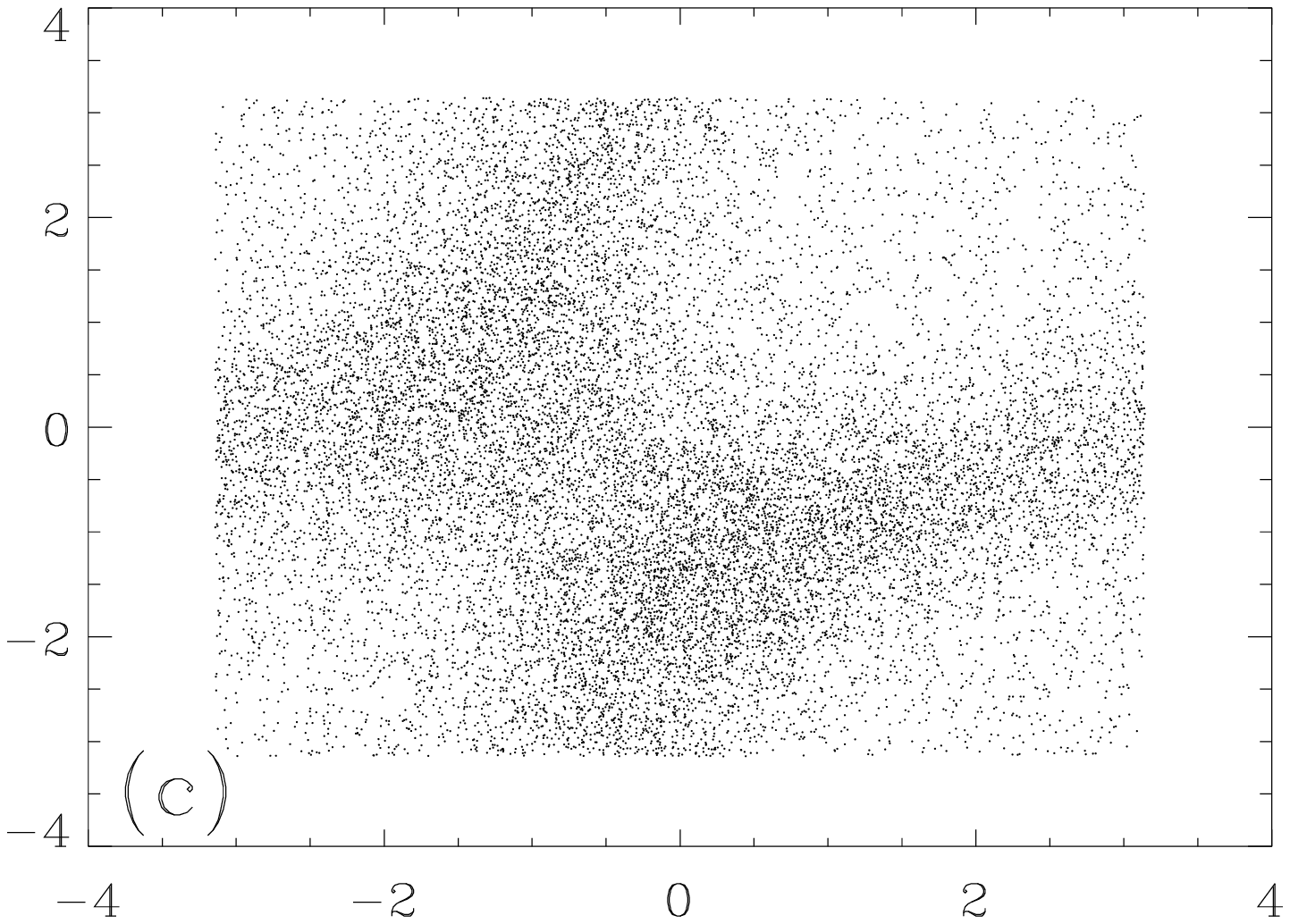}
    \end{minipage}
    \begin{minipage}{4.27cm}
    \epsfxsize=4.27cm
     \epsffile{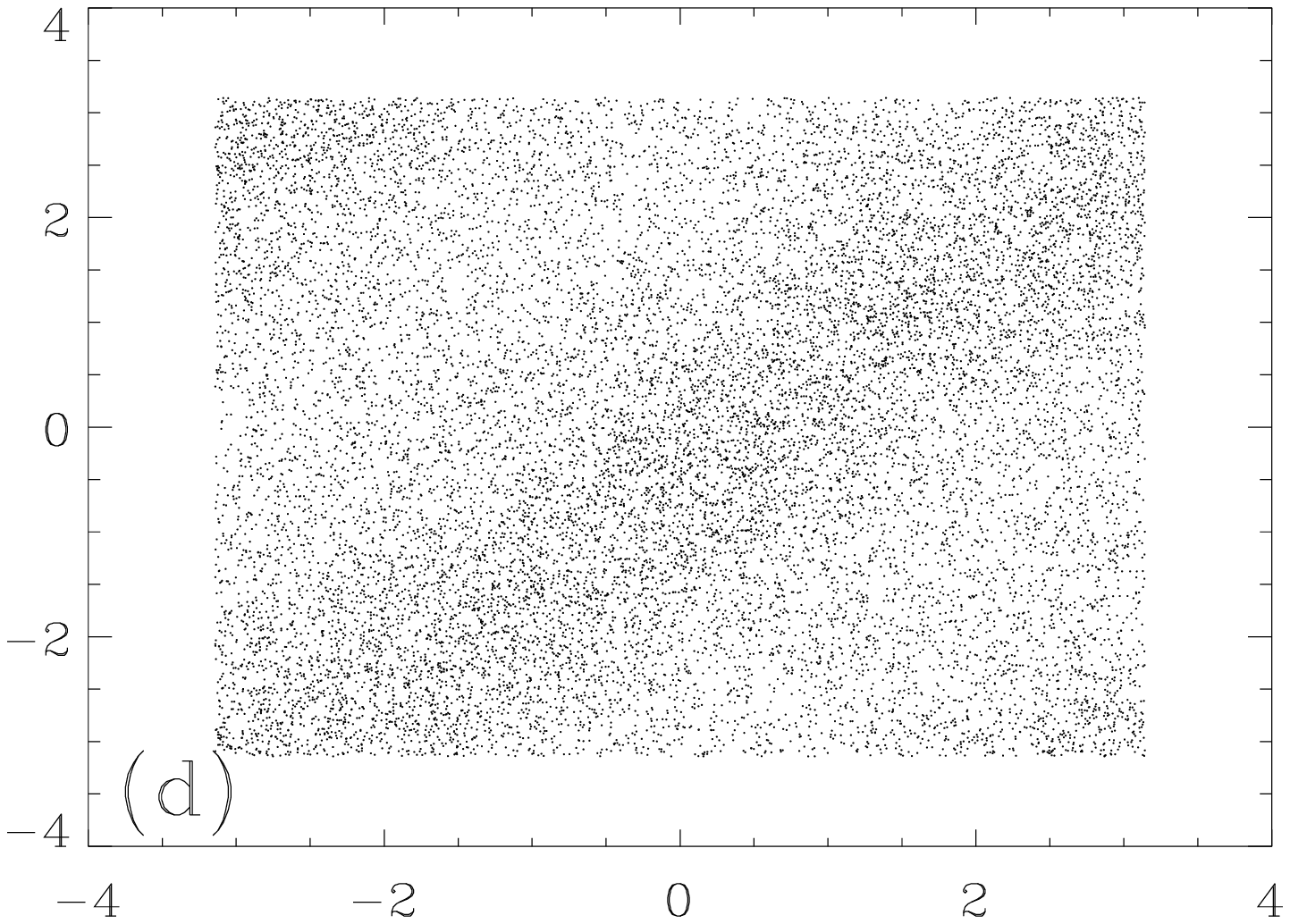}
    \end{minipage}
\end{center}
\caption[fig1]
{ \label{fig1}
a) Time series of the $z$-component of the Lorenz system in a
chaotic regime $\sigma = 10$,  $r = 28$, and $b = 8/3$ and below a surrogate
realization. b) Log-log plot of the scaling of the fluctuations obtained
using the DFA2. The
curves have been shifted to allow for comparison. c)
Phase map of the Lorenz system for $\Delta =
1$. d) Phase map of a surrogate realization for $\Delta = 1$.
}
\end{figure}

\begin{figure}
\begin{center}
\begin{minipage}{3.950cm}
     \epsfxsize=3.90cm
     \epsffile{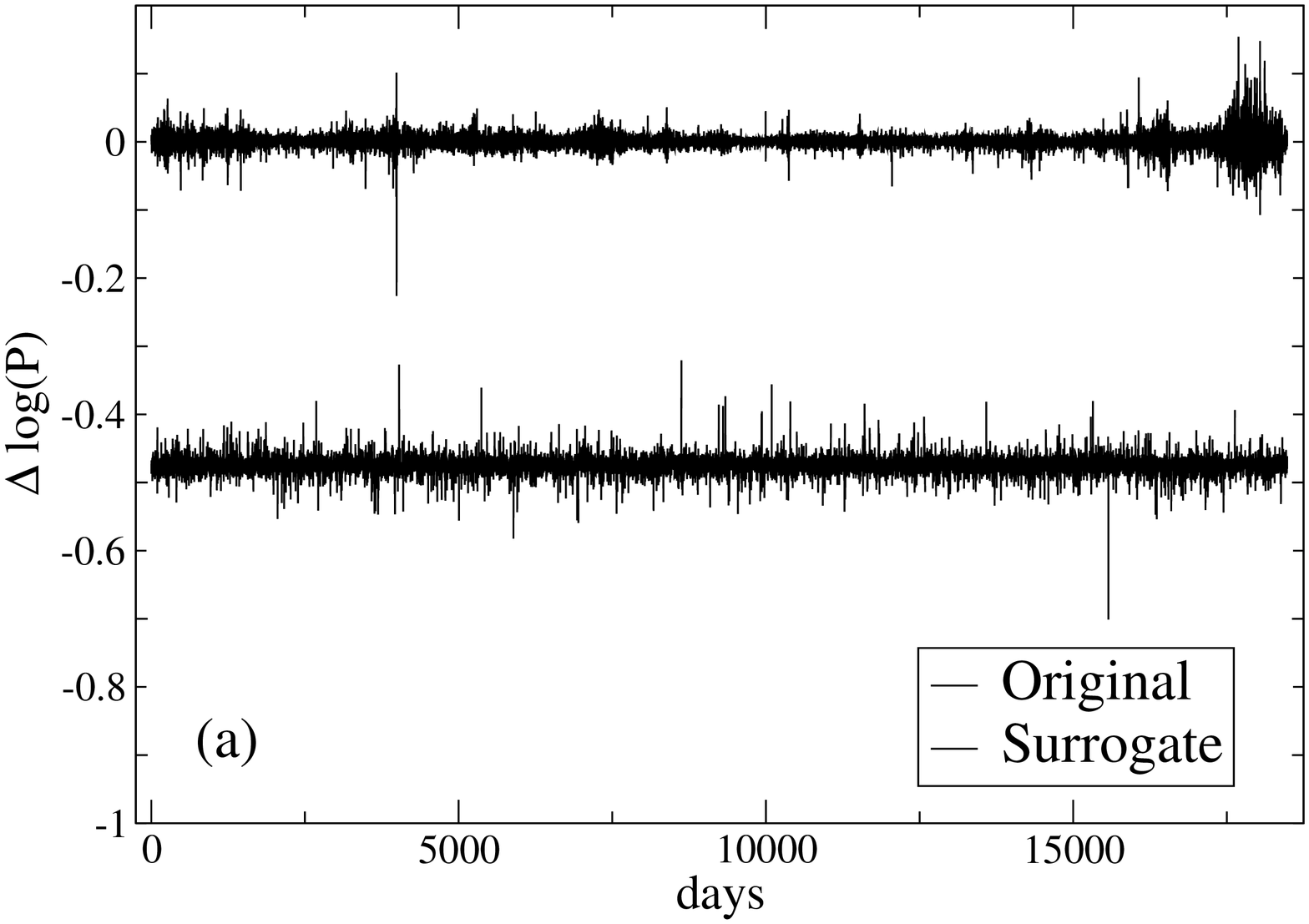}
    \end{minipage}
    \begin{minipage}{3.950cm}
    \epsfxsize=3.90cm
     \epsffile{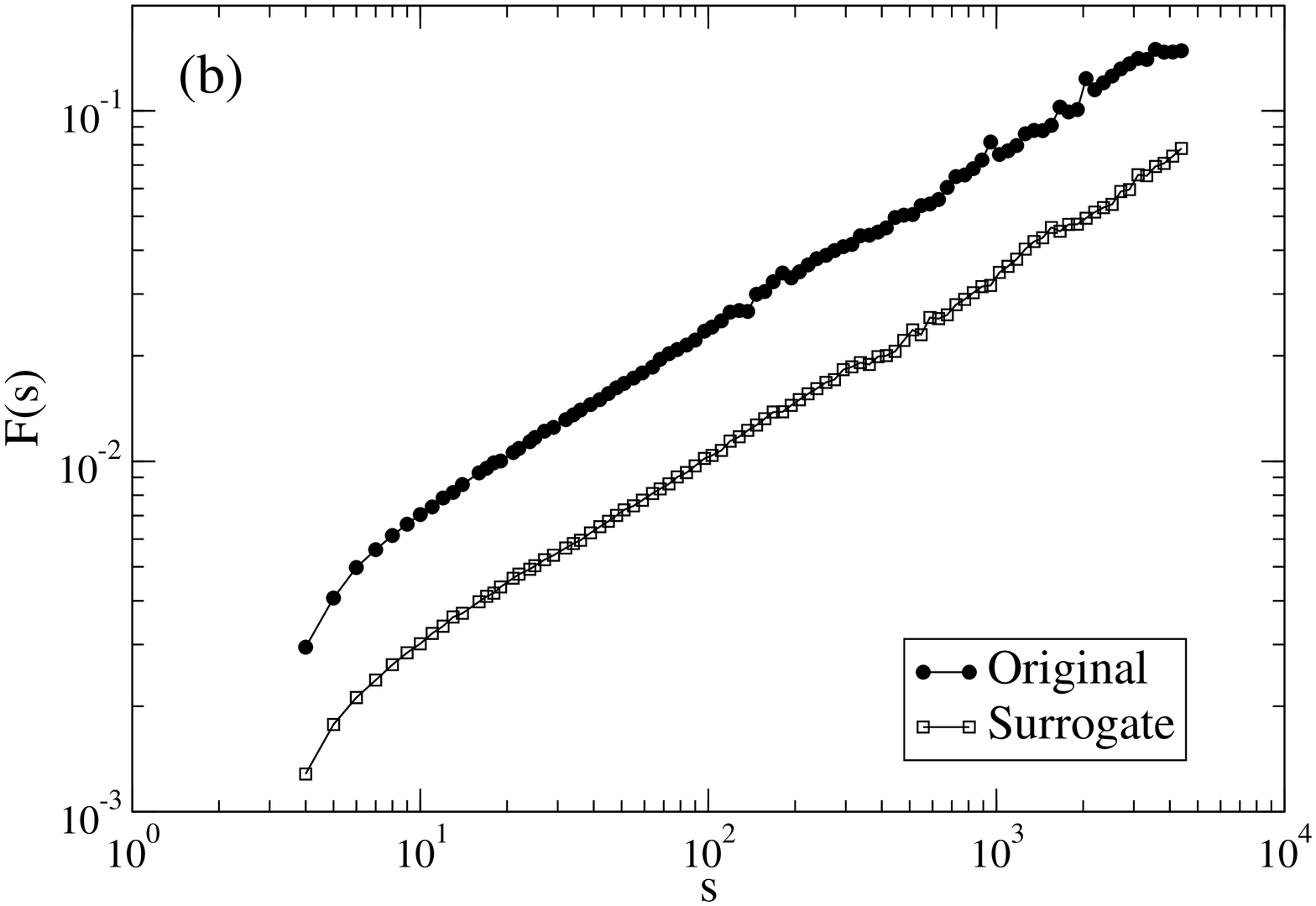}
    \end{minipage}

\begin{minipage}{4.27cm}
     \epsfxsize=4.270cm
     \epsffile{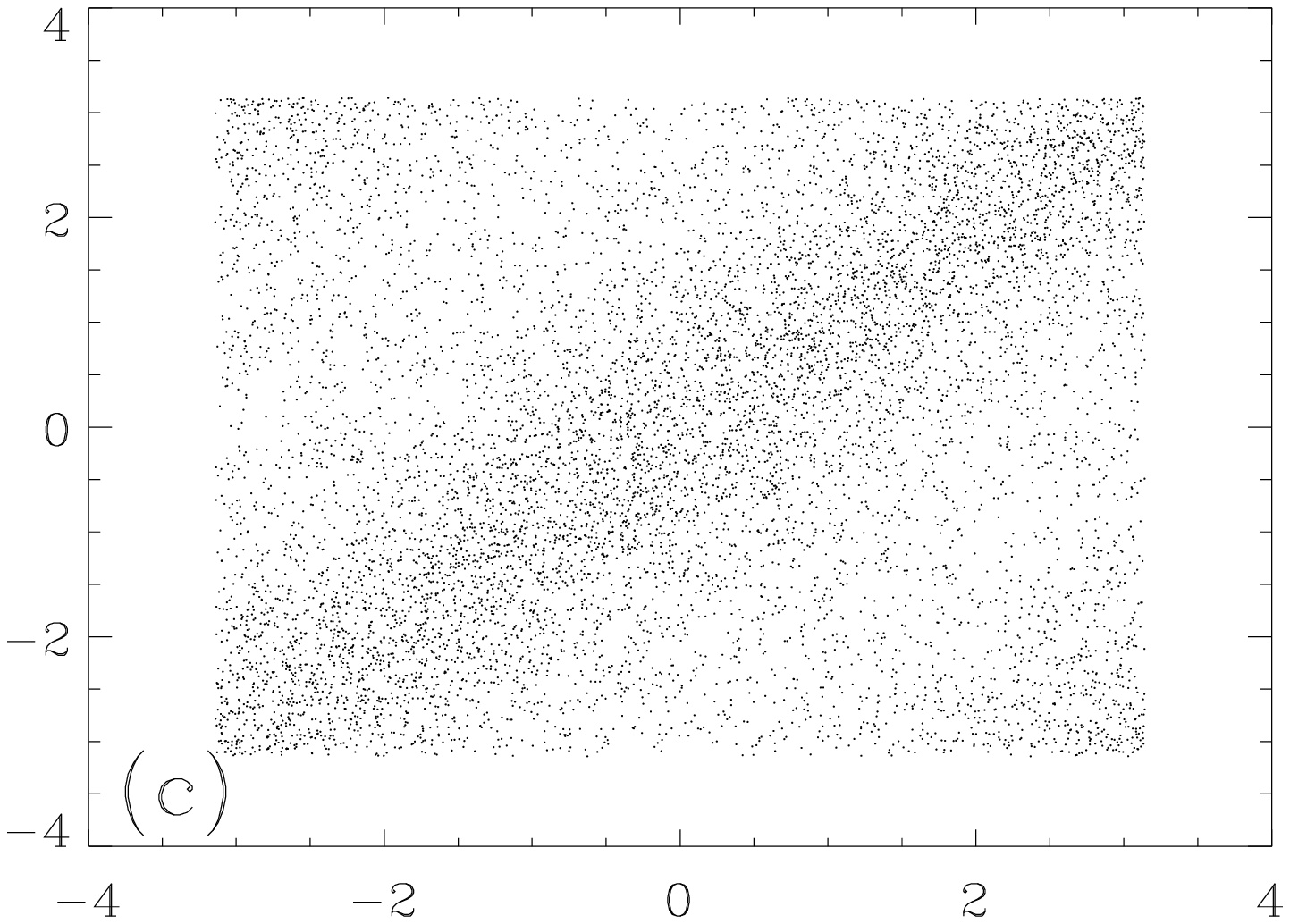}
    \end{minipage}
    \begin{minipage}{4.27cm}
    \epsfxsize=4.270cm
     \epsffile{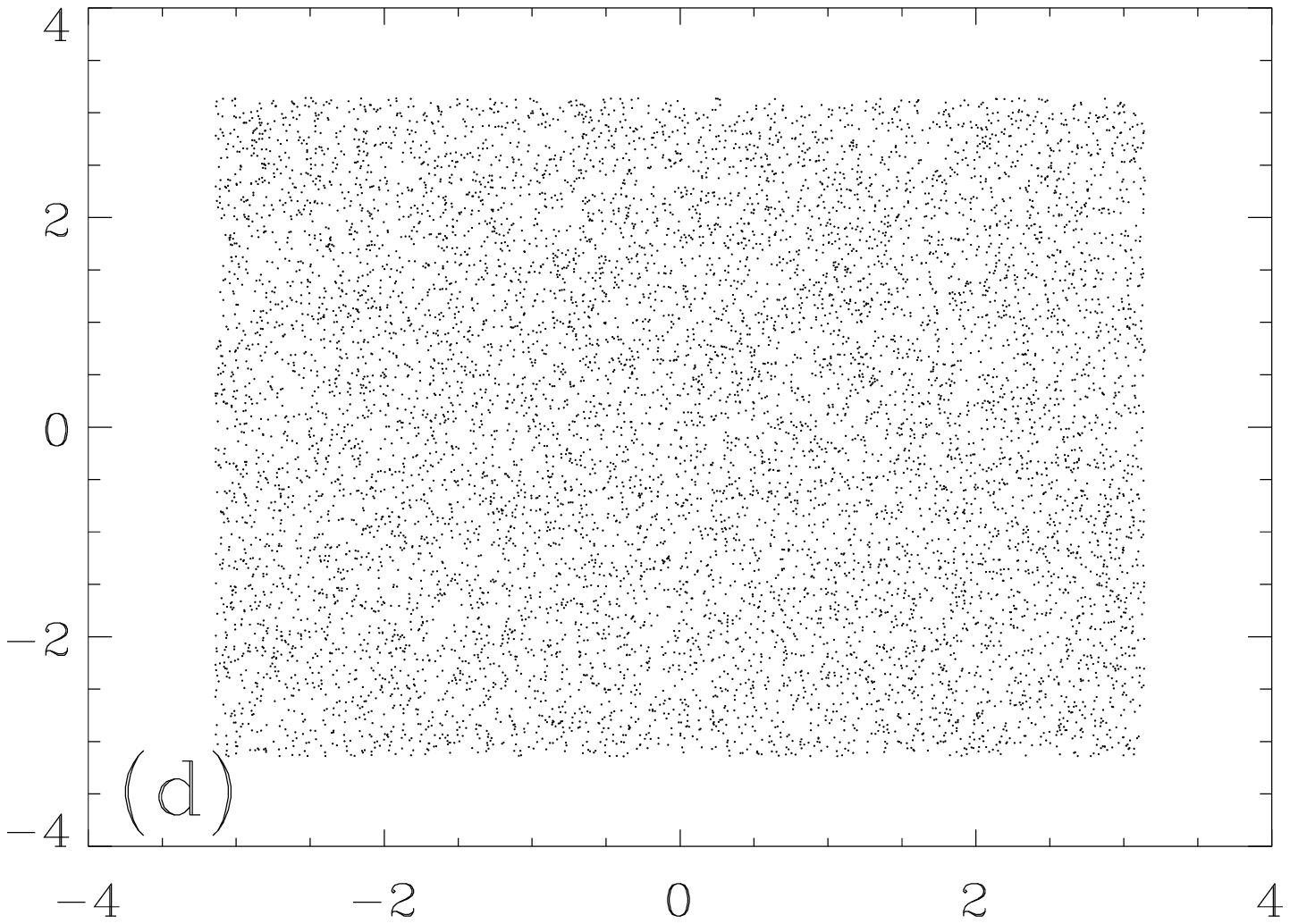}
    \end{minipage}
\end{center}
\caption[fig2]
{ \label{fig2}
a) Time series of the logarithmic daily returns of the Dow-Jones for the
period 1930-2003. Below, a surrogate
realization. b) Log-log plot of the scaling of the fluctuations obtained
using the DFA2. The
curves have been shifted to allow for comparison. c)
Phase map of the logarithmic returns of the Dow Jones for $\Delta =
1$. d) Phase map of a surrogate realization for $\Delta = 1$.
}
\end{figure}
\begin{figure}
\begin{center}
\begin{minipage}{4.25cm}
     \epsfxsize=4.25cm
     \epsffile{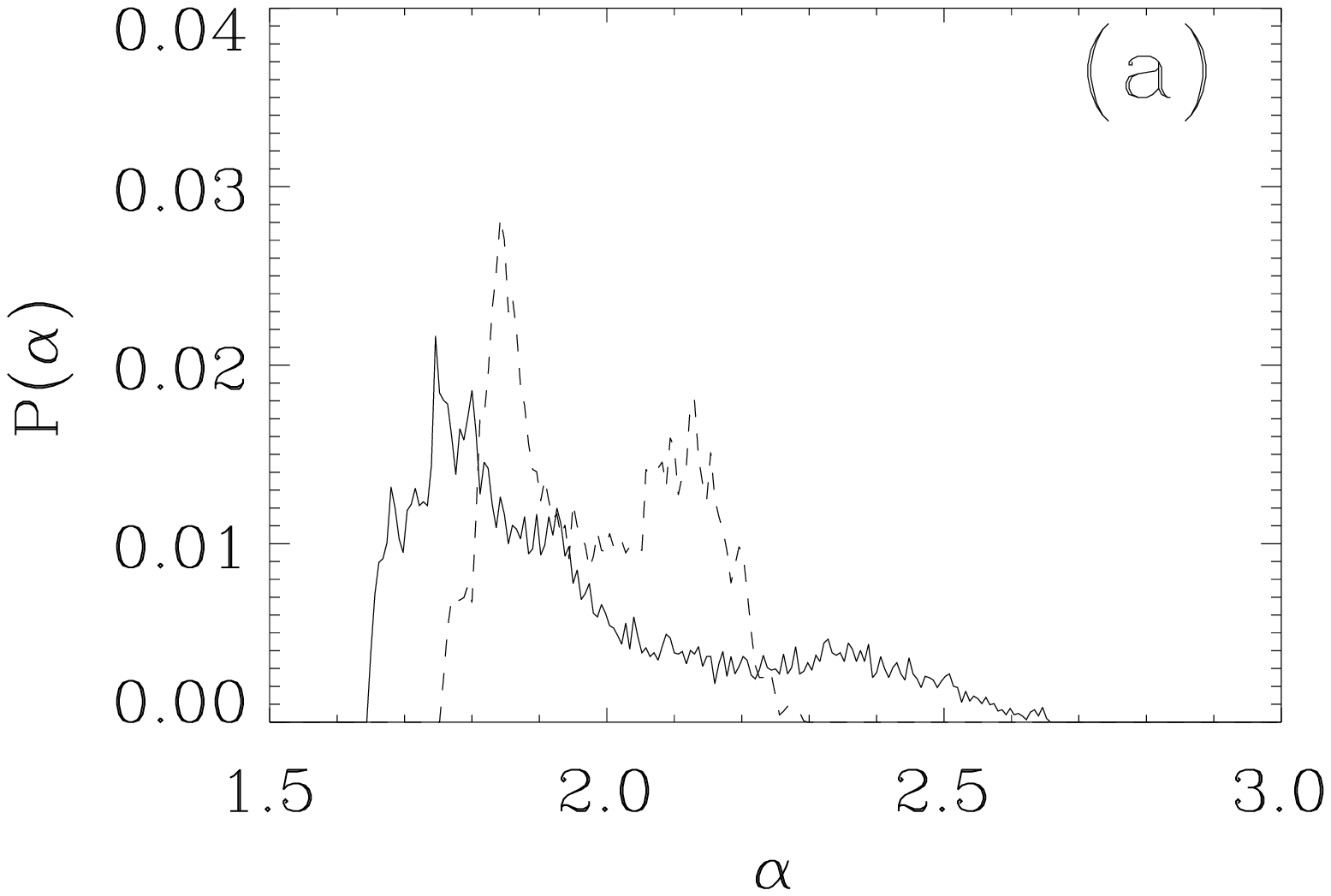}
    \end{minipage}
    \begin{minipage}{4.250cm}
    \epsfxsize=4.25cm
     \epsffile{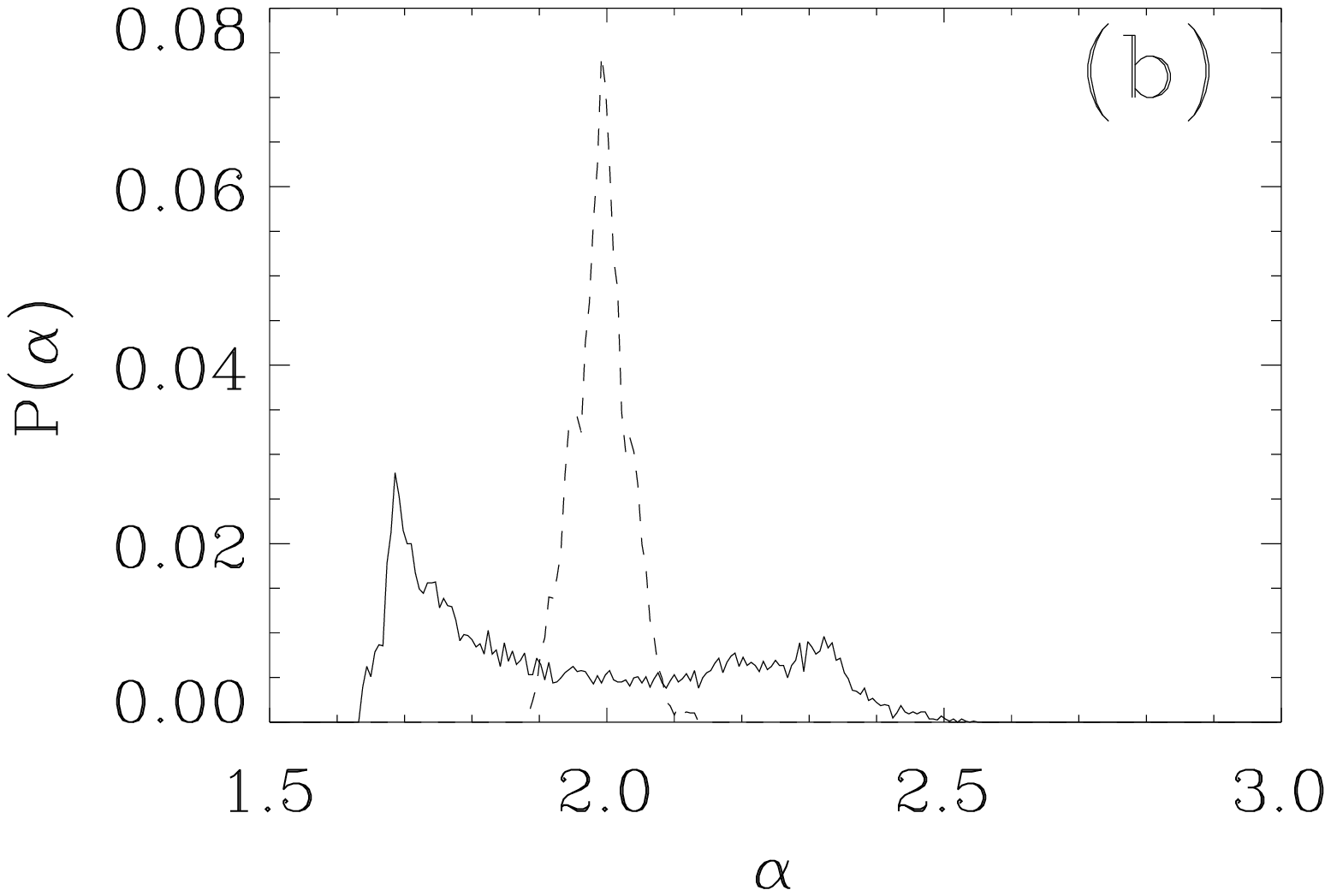}
    \end{minipage}
\end{center}
\caption[fig3]
{ \label{fig3}
Frequency distribution of $\alpha$-values. a) Lorenz
system. b) Dow Jones index. In both cases, the full line (dashed line)
indicates the original phase map (surrogate phase map).}
\end{figure}
%\section{Detecting non-linearities}
Thus, $\Xi$ can be
interpreted as the probability to significantly distinguish between the original
and the surrogate phase maps. The probability $\Xi$ accounts for both the significance
and the extent of the region where a non-uniform structure
exists.

We have applied our statistical test to two time series, namely the
$z$-component of the Lorenz system in a chaotic regime and the
logarithmic stock returns of the Dow Jones. In both cases, we generate
20 surrogates using the ITAAFT algorithm and phase shifts in the range
$\Delta = \{1, \cdots, 100 \}$ are considered. The
phase map structure is analyzed using the SIM with $R=0.8$.

% \section{Applications}
% We have applied our statistical test to two time series, namely the
% $z$-component of the Lorenz system in a chaotic regime and the
% logarithmic stock returns of the Dow Jones. In both cases, we generate
% 20 surrogates using the ITAAFT algorithm and phase shifts in the range
% $\Delta = \{1, \cdots, 30 \}$ are considered. The
% phase map structure is analyzed using the SIM with $R=0.8$.
%\section{The Lorenz system}
Figure 1(a) shows the time series of the $z$-component of
the Lorenz system in a chaotic regime and a surrogate realization. One
can clearly observe differences between the time
series although they have not only the same amplitude distribution but also the
fluctuations show the same scaling behavior (see Fig. 1(b)). The scaling behavior of the
fluctuations was obtained using the Detrended Fluctuation Analysis of
order 2 (DFA2) \cite{10}. This
analysis is equivalent to the power spectrum analysis
\cite{11}. Figure 1(c) and 1(d) show
phase maps for these time series.  We observe that
the phase map of the original time series
displays high and low density regions. Although the phase map of the surrogate time
series is actually more uniform, it still displays a great deal of
structure.  In fact, most of the the surrogate realizations of the
Lorenz system generated using the ITAAFT algorithm display phase
coupling and only some of them are free from phase correlations. 
Figure 3(a) shows the frequency distribution of scaling
indices $P(\alpha)$ for the phase maps shown in  Fig. 1. $P(\alpha)$
of the surrogate realization is narrower and more concentrated around $\alpha = 2$
which is the value expected for a uniform distribution. 
Figure 4(a) shows the significance for a phase
shift $\Delta = 28$. We
observe that a high significance is obtained for a wide range of
scaling index values. The lines at $S = 20$ were included in intervals
where divergencies appear.
Figure 4(c) shows the probability $\Xi$ versus $\Delta$. It indicates
that the probability to significantly differentiate between the
original Lorenz time series and the surrogates oscillates
around $\Xi \sim 0.40$. As
expected, the Lorenz system shows signatures of non-linear
behavior.

Several studies of single stock have focused on the statistical
analyses of the dynamics to model the financial 
markets \cite{19}.
It has been noticed that economic indices exhibit
a non-linear behavior \cite{21,22} which share some qualitative
features with turbulence \cite{23,24,25}. However, understanding the
process that underlies the macroscopic behavior of the stocks remains
at a speculative level. As noticed above, the Fourier
phases are a powerful indicator of the data structure. It is then
relevant to unveil and quantify
the properties of the phases for stock indices. Typically, the
economic indices show a correlated short-time behavior (few days)
which crosses over to an uncorrelated asymptotic behavior. Figure 2(a)
shows the logarithmic returns of the day to day 
closing price of the Dow Jones index in the period 1930-2003 \cite{dj} and a surrogate 
realization. Figure 2(b) shows the scaling behavior of the fluctuations. The
asymptotic regime is governed by an exponent $\gamma \sim 0.5$ which
is the
signature of uncorrelated behavior \cite{10}. However, the phase maps
shown in Fig. 2(c) and 2(d) reveal a new scenario. The phase map for the original
Dow Jones index displays a high density diagonal band and low density
regions. On the other hand, the phase map for the surrogate
realization looks quite uniform. In this case, the ITAAFT algorithm
always generates surrogates free from phase coupling. Figure 3(b)
shows the frequency distribution of scaling indices for the
original and the surrogate phase maps. The
spectrum of scaling indices for the phase
map of the surrogate realization shows a sharp peak
around $\alpha =2$.
\begin{figure}
\begin{center}
\begin{minipage}{4.25cm}
     \epsfxsize=4.25cm
     \epsffile{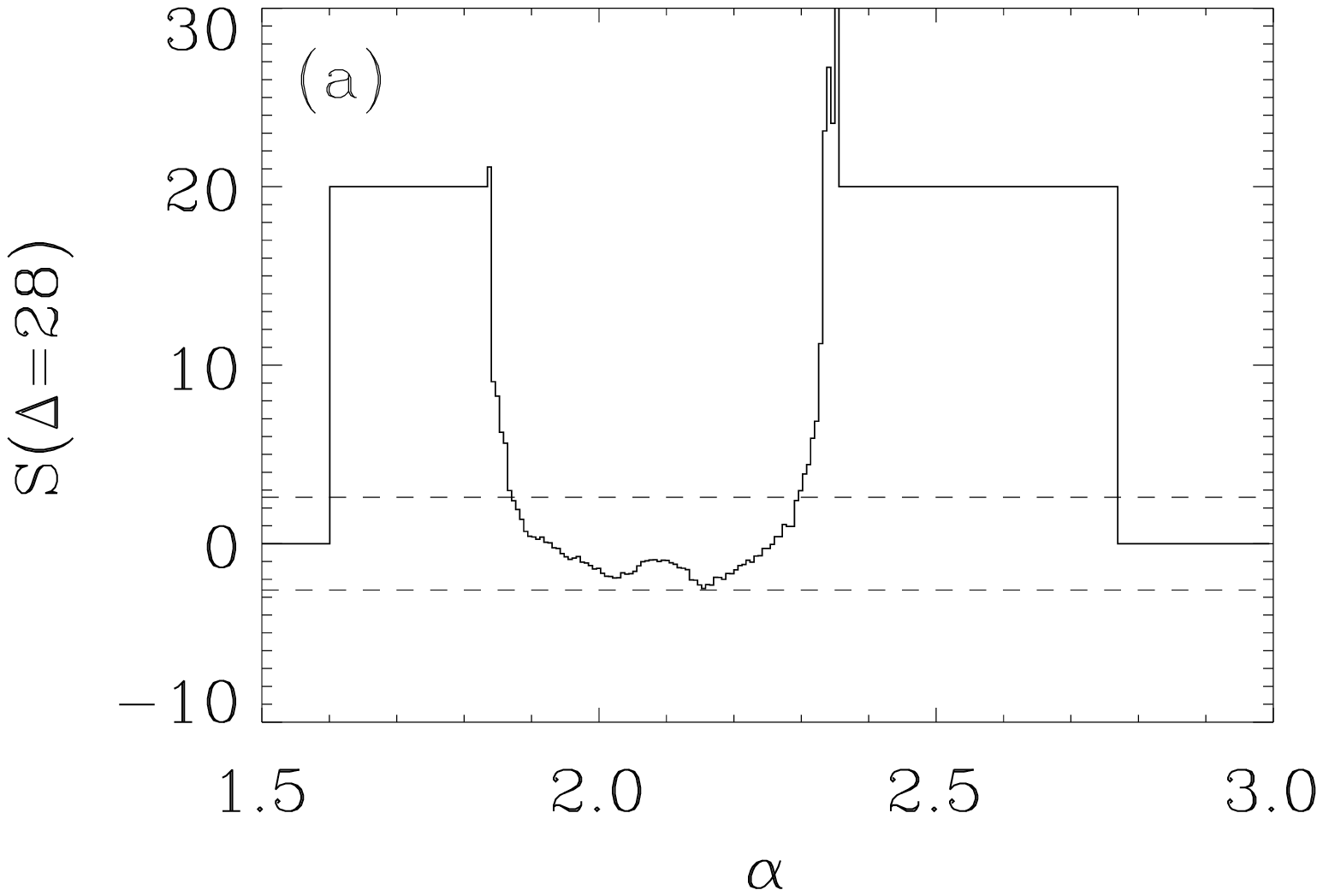}
    \end{minipage}
    \begin{minipage}{4.250cm}
    \epsfxsize=4.25cm
     \epsffile{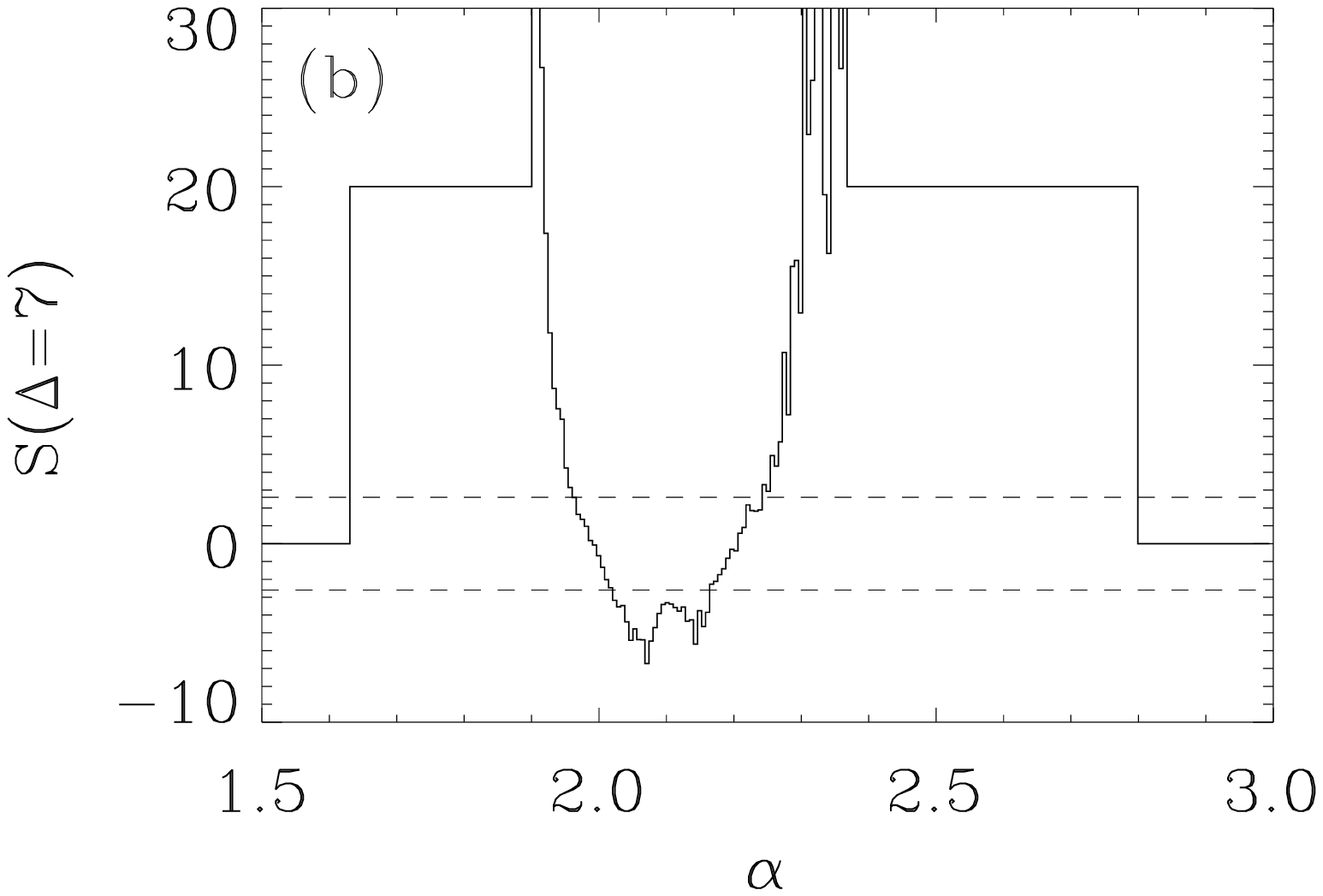}
    \end{minipage}

\begin{minipage}{4.25cm}
     \epsfxsize=4.250cm
     \epsffile{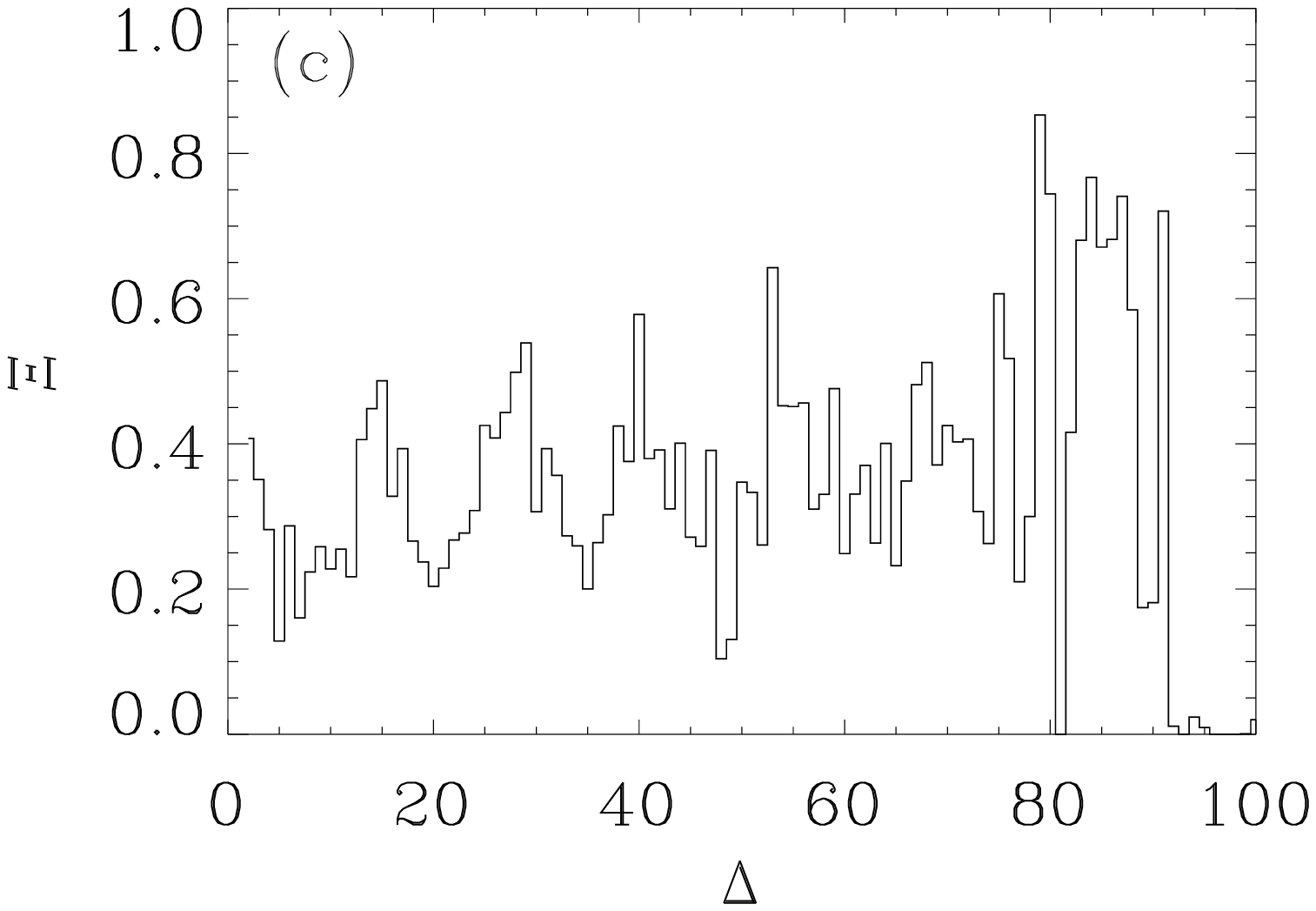}
    \end{minipage}
    \begin{minipage}{4.25cm}
    \epsfxsize=4.250cm
     \epsffile{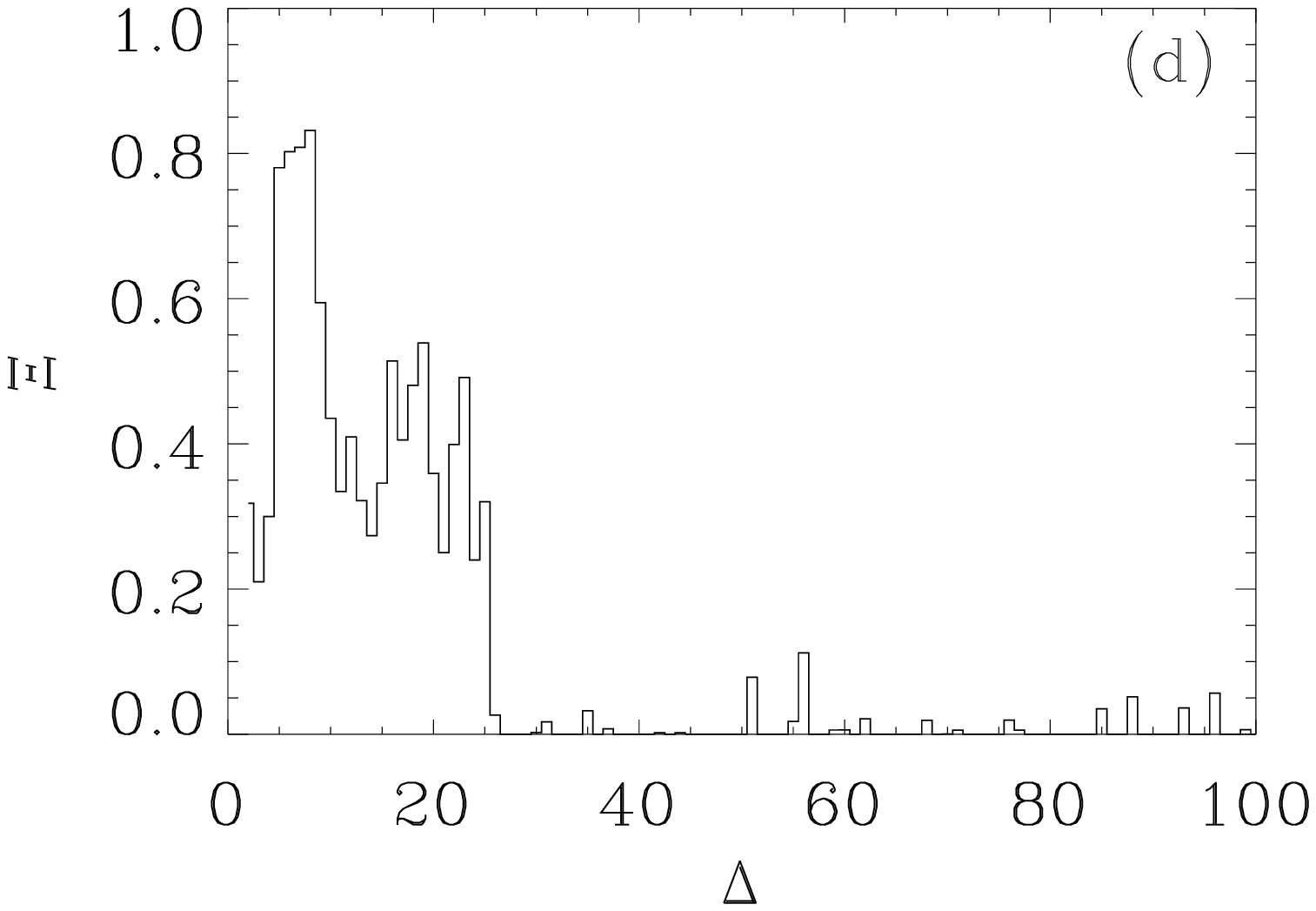}
    \end{minipage}
\end{center}
\caption[fig4]
{ \label{fig4}
a) Significance for Lorenz system. The
dashed horizontal lines at $S = \pm 2.6$ indicate 1$\%$ of the
quantile of the normal distribution. b) Idem for the Dow
Jones index. c) Probability to significantly differentiate between original
and surrogate phase maps for the Lorenz system. d) Idem for the Dow
Jones index.
}
\end{figure}
%\section{The Dow Jones}
As occurred for the Lorenz system, Fig. 4(b) shows that high $S$ values
are obtained for a wide range of
scaling indices. Figure 4(d) indicates 
that there is a high  probability to significantly differentiate between the
original Dow Jones index and the surrogate realizations
for phase shifts up to $\Delta_c \sim 20$. Below
$\Delta_c$, the non-linearities appear to be stronger
than in the Lorenz system. However, increasing the phase shift
$\Delta$ the probability $\Xi$ almost vanishes.

%\section{Conclusions}
We have developed a new test for non-linearities based on the
characterization of the Fourier phase maps using the SIM.
%We have analyzed a prototype non-linear system, namely the
%Lorenz system in a chaotic regime. 
As expected,
the Lorenz system showed signatures of non-linear behavior at all
$\Delta$ scales. However,
we have noted that for this time series the ITAAFT algorithm is 
not always able to generate surrogates free from phase coupling. Then,
our method
could also be used to assess the quality of surrogate data sets since
the performance of the generating algorithms is data dependant. Even
though the fluctuations of the Dow Jones index display an uncorrelated
asymptotic regime, this time series contains non-linearities which seem to be 
stronger than in the Lorenz system.  This is probably because the
ITAAFT algorithm is unable to generate surrogates free from phase
coupling for the Lorenz system. Our results indicate that a novel
characteristic scale of non-linearities $\Delta_c$ exists for the Dow
Jones index.
These findings may be
useful to deeper understand the market dynamics and thus improve the results of risk assessments.

This method can be applied to higher dimensional data sets
since algorithms to generate surrogates in higher dimensions are
available \cite{8,9}.  
In cases where weak non-linearities may be present as in
the Cosmic Microwave Background of radiation \cite{2,3}, the use of local structure measures
for the assessment of
the phase maps, like the scaling indices, instead of global
measures may be more appropriate since global measures may average
important local details of the maps. The development of quite
sensitive tests is particularly relevant for this problem where
the presence or absence of non-Gaussian signatures will support
different evolutionary theories of the universe.
\bibliography{pap_phas_aps.bib}

\end{document}